\documentclass{pasa}%

\renewcommand{\vec}[1]{\mathbf{#1}}

\title[The spheronic universe]
{The spheronic toy universe: how special relativity may be visualized
  to emerge from a wave-nature of matter}
\author[Schmid \& Kroupa]{Manfred
  Schmid$^{1}$\thanks{mschmid@eboracum.de}, 
Pavel Kroupa$^{2}$\thanks{pavel@astro.uni-bonn.de}\\
\affil{
$^{1}$Eboracum GmbH,
Im Vogelsang 9,
53343 Wachtberg, Germany\\
$^{2}$Helmholtz-Institut f\"ur Strahlen und Kernphysik,
  Universit\"at Bonn,  Nussallee 14--16,  53115 Bonn, Germany}}%
\jid{PASA}
\doi{10.1017/pas.\the\year.xxx}
\jyear{\the\year}

\begin{document}%
\begin{abstract}
  We construct an idealized universe for didactic purposes. This universe is
  assumed to consist of absolute Euclidean space and to be filled with
  a classical medium which allows for sound waves. A known solution to
  the wave equation describing the dynamics of the medium is a
  standing spherical wave. Although this is a problem of classical
  mechanics, we demonstrate that the Lorentz transformation is
  required to generate a moving solution from the stationary one. Both
  solutions are here collectively referred to as ``spherons''. These
  spherons exhibit properties which have analogues in the physical
  description of matter with rest mass, among them de Broglie like
  phase waves and at the same time "relativistic" effects such as
  contraction and a speed limit. This leads to a theory of special
  relativity by assuming the point of view of an observer made of such
  spheronic "matter".  The argument made here may thus be useful as a
  visualisation or didactic approach to the real universe, in which
  matter has wave-like properties and obeys the laws of special
  relativity.
\end{abstract}
\maketitle%
\section{INTRODUCTION }
\label{sec:introd}
Relativity and quantum mechanics are the two pillars of modern
physics. Historically, they were discovered separately through the
work of Einstein, de Broglie, Bohr and later Heisenberg,
Schr\"odinger and others. Both theories have so far triumphantly
passed all experimental challenges and provide predictions with
unprecedented precision in their respective field. Both are difficult
to reconcile with our every day experience, that is, classical
physics does not readily allow conceptual access to these two aspects
of our reality.

Here we perform a gedanken experiment. We construct an idealized
universe consisting of a classical medium filling Euclidean
space. This medium will, by construction, allow for the propagation of
classical waves. These can be combined to a particular solution of the
wave equation, which we refer to as a ``spheron''. We show that
spherons which propagate show properties which can be described as
being of quantum mechanical nature, and that they also automatically
obey the laws of special relativity.

This gedanken experiment thus shows how special relativity can be seen
as possibly being a direct consequence of the wave nature of matter, but
perhaps more importantly, it may be a useful didactic argument for
introducing the concepts of quantum mechanics and special relativity.

The structure of this contribution is as follows: Sec.~\ref{sec:Lor}
introduces the idealized universe and the ``matter'' which can exist
in it. The properties of these ``matter particles'', which are
propagating standing waves and are referred to as spherons, are
discussed in Sec.~\ref{sec:spherons}. Here it is shown that these
spherons have quantum-mechanical-like properties, and special
relativistic behavior emerges naturally. A discussion of these issues
is provided in Sec.~\ref{sec:disc}, where the insights gained are
suggested to be potentially helpful in understanding or visualizing
the possibly deep and natural connection between the wave nature of
matter and the emergence of special relativity in a universe where
matter has wave-like properties. Appendix~A compares spherons in the
idealized universe with de~Broglie waves in the real universe, and
Appendix~B contains an in-depth treatment of operator methods to be
used on analyzing spherons in the idealized universe. In Appendix~C a
relevant and well-motivated question is asked, namely if the motion of
spherons relative to the ideal gas may be detectable from the inside
by a ``spheronic observer'', therewith demonstrating the
spheronic-universe ansatz to not lead to special relativistic
behavior. This question is analysed in depth and explicitly
computed. The result is that in an idealized universe consisting only
of spherons such that all measurements can only be performed by
internal observers made of spherons with measuring devices made of
spherons, the existence of the ideal gas cannot be inferred by a wind
or from an anisotropic sound speed.  Possible implications for the
real universe are discussed.

\section{THE IDEALIZED UNIVERSE}
\label{sec:Lor}
\begin{figure}
\includegraphics[width=8cm]{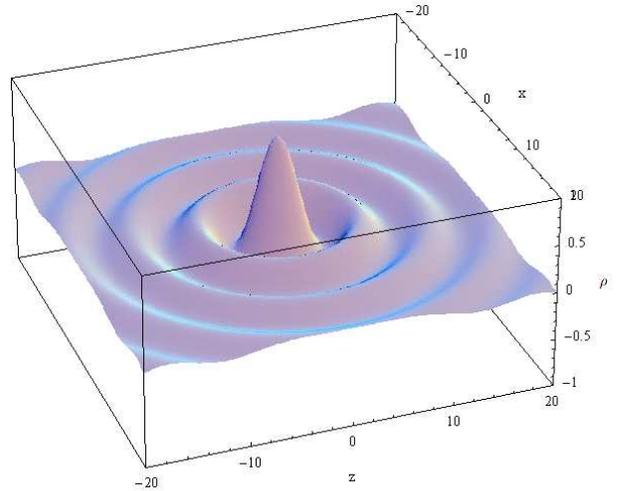}
\caption{$\rho$ from eq.~\ref{eq:stat-spheron-expl} with $A_{0}=\frac{1}{2}$,
$\omega=1$ and $c_s=1$ at $t=0$ (units are arbitrary).} 
\label{fig:spheron_stat}
\end{figure}

Our idealized classical Euclidean universe we consider, for the sake of the
argument, to be filled with an idealized gas. Such a gas can only
harbor curl-free oscillations which may be described by a scalar
quantity (e.g. the time-varying density difference from the ambient
medium, $\rho(t)$). The propagation of small perturbations in the
density of the medium with sound speed $c_s$ are solutions of the wave
equation:
\begin{equation}
\frac{1}{c_s^{2}}\frac{\partial^{2}}{\partial t^{2}}\rho-\nabla^{2}\rho=0.
\label{eq:3d-wave-equation}
\end{equation}
For a derivation of the wave equation from the properties of the
medium and a consideration of the limitations of the linear approach,
see e.g. the textbook by Skudrzyk (1971). Since our considerations
only rely on the wave equation describing the dynamics of the medium
sufficiently accurately, we do not discuss this topic here.  A class
of solutions of the wave equation are plane waves,
\begin{equation}
\rho\left(x,y,z,t\right)=A_{0}{\rm sin}\left(kx-\omega t\right).
\label{eq:propagating-plain-wave}
\end{equation}
The frequency, $\omega$, and wavenumber, $k$, are
coupled by $\omega=c_s k$ which holds for all solutions presented in
this contribution. Since the wave equation is linear, plane waves may be
superimposed. Superimposing the above wave with an identical wave
moving in the opposite direction results in a standing wave,
\begin{eqnarray}
\rho &= &A_{0}{\rm sin}\left(kx-\omega t\right)+A_{0}{\rm sin}\left(kx+\omega
  t\right)\nonumber \\ 
&= &2A_{0}{\rm sin}\left(kx\right){\rm cos}\left(\omega t\right).
\label{eq:standing-plain-wave}
\end{eqnarray}
Formulated in the usual spherical coordinates with radial distance
$r=\left(x^{2}+y^{2}+z^{2}\right)^{\frac{1}{2}}$ and Cartesian
coordinates $x$, $y$, $z$,
\begin{equation}
\rho=\frac{A_{0}}{r}{\rm sin}\left(kr-\omega t\right)
\label{eq:ougoing-sph-w}
\end{equation}
is a solution of the wave equation (eq.~\ref{eq:3d-wave-equation}) and
describes an outbound spherical wave. Here, the wave ``starts'' from a
point source at the origin of the coordinate system and wave fronts
subsequently propagate with speed $c_s$ in direction of increasing
$r$. Superposition with an inbound wave leads to
\begin{equation}
  \rho = 2A_{0}\frac{{\rm sin}\left(kr\right)}{r}{\rm cos}\left(\omega t\right),
\label{eq:stat-spheron-expl}
\end{equation}
which is a standing spherical wave. Its two parts are the
amplitude $A\left(r\right)=2A_{0}{\rm sin}\left(kr\right)/r$, which only
depends on position, and a harmonic oscillation $B\left(t\right)={\rm cos}\left(\omega t\right)$,
which only depends on time. Thus 
\begin{equation}
\rho=A\left(r\right)B\left(t\right).
\label{eq:stat-spheron}
\end{equation}
Choosing a plane (here the x-z plane), the value\\
$\rho\left(x, y=0, z, t=0\right)$
is plotted on the vertical-axis in Fig.~\ref{fig:spheron_stat}. 
%-------------------FIGURE----------------------
%-------------------FIGURE----------------------

Standing spherical waves do have a notion of localization. The point
$\vec{o}$ where the amplitude $A\left(r\right)$ reaches its
global maximum is called the wave center, which is in the case of the
above example the origin. It is one of the key parameters for a description
of the standing spherical wave, complemented by the frequency $\omega$.
Since there is only one single frequency $\omega$ involved, this
is a monochromatic phenomenon.

With the speed of sound, $c_s$, being the limiting speed of the medium, and setting
\begin{eqnarray}
\gamma_s & = & \sqrt{1-\frac{v^{2}}{c_s^{2}}},\label{eq:gamma}\\
t' & = & \frac{1}{\gamma_s}(t-\frac{v}{c_s^{2}}z),\\
z' & = & \frac{1}{\gamma_s}(z-vt),\label{eq:zs}\\
r' & = & \left(x^{2}+y^{2}+z'^2\right)^{\frac{1}{2}},\label{eq:rs}
\end{eqnarray}
then, provided $v^{2}<c_s^{2}$,
\begin{eqnarray}
\rho & = & 2A_{0}\frac{{\rm sin} (kr')}{r'}{\rm cos} (\omega t'),\label{eq:mov-spheron-expl}\\
 & = & A(r') B(t')\label{eq:mov-spheron}
\end{eqnarray}
also solves the wave equation. For the velocity $v\rightarrow0$,
eq.~\ref{eq:mov-spheron} smoothly turns into eq.~\ref{eq:stat-spheron}
so the former may be seen as the latter in propagation along the
$z$-axis. From here on we will refer to both of them as a ``spheron'',
which may, accordingly, stay put or propagate. Note that
$\rho(x',y',z',t')$ given by eq.~\ref{eq:mov-spheron} is a spheron in
motion as is signified by the primed coordinates.

Equations ~\ref{eq:gamma} - ~\ref{eq:zs} are close to the formulas for
the Lorentz transformation used in relativity and the use of these
formulas within classical mechanics is unusual. We note explicitly
here that this has nothing to do with the theory of special
relativity, but merely constitutes the mathematical description of
propagating spherons, which are physical solutions of the classical
wave equation. An explicit calculation quickly shows that
eq.~\ref{eq:mov-spheron} solves the wave equation whereas the use of
quantities resulting from a Galilean transformation in
eq.~\ref{eq:stat-spheron} does not result in such a
solution. Mathematically this will not come as a surprise, since the
wave equation is known as an invariant of the Lorentz transformation
and it is not an invariant of the Galilean transformation.

Equations ~\ref{eq:gamma} - ~\ref{eq:zs} should therefore be thought
of as a "Lorentz composition of quantities" instead of a Lorentz
transformation in order to avoid confusion with relativistic ideas and
terminology. So while the use of such a Lorentz composition is
commanded by the wave-nature of spherons, we have defined our factor
$\gamma_s$ not in the usual way used in special relativity (the usual
notation applied there is $\gamma=1/\gamma_s$) to remind us of the
fact that we are not explicitly dealing with special relativity
here. Instead, the idealized universe is based on classical mechanics
in classical space and time.

We now populate this universe with matter. To represent matter without
rest mass (radiation) we choose moving plane waves that follow
eq. ~\ref{eq:propagating-plain-wave}. Such waves transport energy at
the speed of sound and cannot stand still. We choose spherons to
represent matter with rest mass. Our choice is justified by the
similarities between the properties of spherons in the idealized universe
and the properties of matter with rest mass in our real universe,
which are laid out in the next section. From these elementary
particles we envision more complex material entities being built,
including intelligent beings that are able to observe their
environment and perform measurements with their material tools, all of
which are composed of elementary particles which are spherons in the
idealized universe. We call such an intelligent being a "spheronic
observer".

For the sake of the present argument we ignore the fact that the
spherons cannot interact. This is due to the linearity which is a
consequence of the assumptions made and is addressed later. We
reiterate that the idealized universe is filled with spherons which are its
elementary particles with rest mass. The gas is merely a
(hypothetical) medium which aids our discussion and which sustains the
existence of the spherons. Observers made of spheronic ``matter''
cannot detect this gas as explained in the next section. 

\section{SPHERONS}
\label{sec:spherons}
The wave center of eq.~\ref{eq:mov-spheron} is moving with speed $v$
in the $z$-direction as can be seen by setting $z'=0$ and solving for
$z$. So compared to a standing spherical wave, a spheron naturally
needs the velocity vector $\vec{v}$ as an additional parameter for its
description. Eq.~\ref{eq:mov-spheron} also shows that a spheron may be
understood to be composed of two parts: The amplitude $A(r^{'})$ and
an oscillating part $B(t^{'})$.

From the transformation it is evident that the speed of sound is the
limiting speed for the spheron. Physically, this will come as no
surprise, since the speed of sound is the limiting speed of energy
transport in the medium and a propagating spheron is ``transporting''
energy.

Note that an observer consisting of spherons does not experience a
resistance from the medium while moving. There is no "airflow" around
a spheron because spherons propagate as waves and are not solid bodies
moving through the medium.

\subsection{Length contraction and time dilation}
\label{sec:contdil}

A spheronic observer has no ab-initio knowledge of space-time and
has to derive measurement rods and clocks from the properties of
spheronic matter.

To define the rods of a comoving spheronic observer, let her or his 
defining standard ruler stretch from the wave center to the first zero
of the amplitude $A(r^{'})$. While this delivers the desired length
scales derived from the properties of a spheron, the standard ruler
should not be thought of as a thing made of ``solid real'' matter. It
remains a wave-natured measurement tool and for a spheronic observer a
measurement is a process determined by the properties of her or his 
tools

Every existing real clock relies on the counting of some cyclic event
provided by the matter the clock is made of. To define a spheronic
clock, let the observer count the oscillations in the wave center. This
wrist watch then defines her/his proper time.

\begin{figure}
  \centering{}\includegraphics[width=8cm]{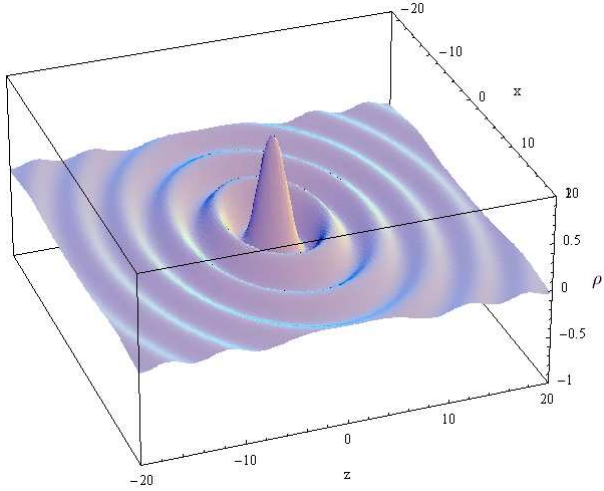}\caption{As
    Figure~\ref{fig:spheron_stat}. Amplitude of $\rho$ from
    eq.~\ref{eq:mov-spheron} with $A_{0}=\frac{1}{2}$, $\omega=1$,
    $c_s=1$ and $v=0.75c_s$.} \label{fig:spheron_mov}
\end{figure}

The effect of the Lorentz-composition on the amplitude $A(r^{'})$
is shown in Fig.~\ref{fig:spheron_mov}. Compared to
Fig.~\ref{fig:spheron_stat}, the waves centered around the wave center
in Fig.~\ref{fig:spheron_stat} have acquired an elliptic shape. This
is a ``relativistic'' contraction in direction of propagation.  The
quotes are meant to indicate once more that this ``relativistic''
contraction is based on the speed of sound as the limiting speed, and
not on the speed of light in our real universe. From here on, we
assume that the reader keeps this in mind and leave the quotes away
for ease of reading. Besides this difference in limiting speeds, the
effect follows the pattern of a contraction in special relativity in
the real universe, which is a consequence of the formulas used.

Writing out $B(t') = {\rm cos} (\omega \, t')$
\begin{eqnarray}
B(t) & = & {\rm
  cos}\left(\frac{\omega}{\gamma_s}\left(t-\frac{v}{c_s^{2}}z\right)\right),
\label{eq:real-de-Broglie-wave}
\end{eqnarray}
and for a virtual (spheronic) observer at rest, say at $z=0$, the frequency becomes
\begin{equation}
B\left(t^{'}\left(z=0\right)\right)={\rm cos}\left(\frac{\omega}{\gamma_s}\left(t\right)\right), \label{eq:toph-1}
\end{equation}
and thus $\omega'=\omega/\gamma_s>\omega$. The term
\begin{equation}
\omega'=\frac{\omega}{\sqrt{1-\frac{v^{2}}{c_s^{2}}}},
\label{eq:w-rel}
\end{equation}
resembles, in the real universe, the mass term, 
\begin{equation}
m=\frac{m_{0}}{\sqrt{1-\frac{v^{2}}{c^{2}}}},
\end{equation}
of special relativity, where $m_{0}$ is the rest mass. 

In the idealized universe, for a virtual spheronic observer comoving with
the wave center, $z=vt$, and thus the oscillation is
\begin{eqnarray}
B\left(t^{'}\left(z=vt\right)\right) & = & {\rm cos}\left(\frac{\omega}{\gamma_s}\left(t-\frac{v}{c^{2}}vt\right)\right),\\
 & = & {\rm cos}\left(\gamma_s\omega t\right),\label{eq:toph-2}
\end{eqnarray}
where $\gamma_s\omega<\omega$ may be interpreted as a dilation of proper time. 

%-------------------FIGURE----------------------
%-------------------FIGURE----------------------

\subsection{Phase waves and operator methods}
\label{sec:OpMeth}

The wave part of the spheron at rest is a harmonic oscillation.  The
wave part of a propagating spheron (eq.~\ref{eq:real-de-Broglie-wave})
is a moving plane wave with phase velocity
\begin{equation}
v_{P}=\frac{c_s}{v}c_s,
\end{equation}
as can be seen from setting $t-vz/c_s^{2}=0$ and solving for $z$.  The
speed of the phase is larger by a factor $c_s/v$ than the speed of
sound. This supersonic phase speed resembles the superluminal phase
speed of a quantum mechanical de Broglie wave. The phase speed does
not interfere with the speed of sound as the limiting speed of energy
transport in the medium, since this phase wave does not transport any
energy. The same applies to a de Broglie wave. Also
eqs.~\ref{eq:toph-1} and~\ref{eq:toph-2} may be seen as a ``theorem of
phase harmony'' (``th\'eor\`eme de l'harmonie de phases'', de Broglie
1925) in action.

\begin{figure}
\includegraphics[width=8cm]{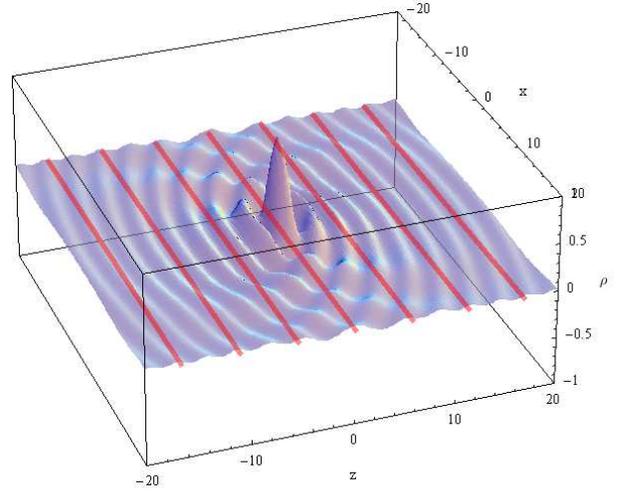}
\caption{As Figure~\ref{fig:spheron_mov}. Spheron with
    $A_{0}=\frac{1}{2}$, $\omega=1,$ $c_s=1$ and $v=0.75 c_s$. The
    wavefronts of the phase wave are marked in red.}
\label{fig:amplitude_prop}
\end{figure}

The plane wave described by eq.~\ref{eq:real-de-Broglie-wave} is in
many respects a ``phase wave'' as is the original ``onde de phase'' of
de Broglie and could therefore be termed a "real de Broglie wave".
Fig.~\ref{fig:amplitude_prop} shows a moving spheron as the multiplication of a contracted
amplitude with a plane phase wave.
%-------------------FIGURE----------------------
%-------------------FIGURE----------------------

It is common practice to employ complex quantities (see e.g. Skudrzyk,
1971) for the description of plane waves, since that simplifies the
calculations considerably. The phase wave of a propagating spheron may
be written in complex notation as $e^{i\omega t'}$.  For velocities
small compared to the speed of sound, $v\ll c_s$, an approximation may
be used. Using the identities
\begin{eqnarray}
\frac{\partial}{\partial t}e^{i\omega t'} & \equiv & i\frac{\omega}{\gamma_s}e^{i\omega t'},\\
\nabla^{2}e^{i\omega t'} & \equiv & \left(i\frac{\omega}{\gamma_s}\frac{\vec{v}}{c_s^{2}}\right)^{2}e^{i\omega t'},
\end{eqnarray}
and a Taylor expansion of $\omega/\gamma_s$ around $v=0$ where the third and higher powers of $v$ are treated as too small to count, then
\begin{equation}
\frac{\omega}{\gamma_s}\approx\omega\left(1+\frac{1}{2}\frac{v^{2}}{c_s^{2}}\right),
\end{equation}
and it follows that 
\begin{eqnarray}
\frac {\partial} {\partial t} e^{i\omega t'} & \approx & i\omega\left(1+\frac{1}{2}\frac{v^{2}}{c_s^{2}}\right)e^{i\omega t'},\\
\nabla^{2}e^{i\omega t'} & \approx & -\omega^{2}\frac{v^{2}}{c_s^{4}}e^{i\omega t'},
\end{eqnarray}
from which 
\begin{eqnarray}
-i\frac {\partial} {\partial t}  e^{i\omega t'}  &= & \omega e^{i\omega t'}-\frac{c_s^{2}\nabla^{2}e^{i\omega t'}}{2\omega}
\end{eqnarray}
can be concluded. Remarkably, this resembles a
``Schr\"odinger-Equation'' for an unbound particle. Slowly ($v\ll
c_s$) moving elementary particles of the idealized universe, slow spherons,
thus obey a Schr\"odinger-like equation.  Also, for purely algebraical
reasons, this phase wave is a solution of a ``Klein-Gordon-Equation'',
where the frequency $\omega$ plays the role of the mass in the real
Klein-Gordon equation,
\begin{equation}
\left(\frac{1}{c_s^{2}}\frac{\partial^{2}}{\partial t^{2}}-\nabla^{2}\right)e^{i\omega t'}=-\frac{\omega^{2}}{c_s^{2}}e^{i\omega t'}.
\end{equation}
All this makes it tempting to explore the applicability of other quantum mechanical mathematics to spherons. 

Written in complex notation, eq.~\ref{eq:mov-spheron-expl} becomes
\begin{equation}
\rho=2A_{0}\frac{{\rm sin}\left(kr'\right)}{r'}e^{i\omega t'},
\label{eq:complex-mov-spheron}
\end{equation}
and
\begin{equation}
\rho^{*}=2A_{0}\frac{{\rm sin}\left(kr'\right)}{r'}e^{-i\omega t'}
\end{equation}
is its complex conjugate. Accordingly,
\begin{equation}
\rho\rho^{*}=\left(2A_{0}\frac{{\rm sin}\left(kr'\right)}{r'}\right)^{2},
\end{equation}
since $e^{i\omega t'}e^{-i\omega t'}=1$ always holds. The volume
integral over $\rho\rho^{*}$ is
\begin{eqnarray}
\int\rho\rho^{*}dV & = & 4A_{0}^{2}2\pi2\gamma_s\left[\frac{1}{2}r'-\frac{1}{4k}{\rm sin}\left(2kr'\right)\right]_{0}^{R'}
\end{eqnarray}
(see Appendix~B), where $R'$ is the radius associated with the volume
$V$. The integral is not a constant value. To enforce a value of~$1$
for the volume integral, a factor
\begin{equation}
\vartheta'=\frac{1}{2A_{0}}\frac{1}{\pi r'}\sqrt{\frac{1}{2k\gamma_s}}
\end{equation}
 may be used (see Appendix~B). Setting
\begin{eqnarray}
\chi & = & \vartheta' 2A_{0}\frac{{\rm sin}\left(kr'\right)}{r'},\\
 & = & \frac{1}{\pi r'}\sqrt{\frac{1}{2k\gamma_s}}\frac{{\rm sin}\left(kr'\right)}{r'},
\end{eqnarray}
and 
\begin{equation}
\psi=\chi e^{i\omega t},\label{eq:sp-qm-representation}
\end{equation}
leads to 
\begin{equation}
\int\psi^{*}\psi dV=\int\chi^{2}dV=1.
\end{equation}
The quantity $\psi$ may be seen as representing physical quantities of
eq.~\ref{eq:mov-spheron}, since it allows for a recovery of elementary
properties of the spheron with operator-methods in the following way
(for a detailed calculation, see Appendix~B): Define
\begin{equation}
\vec{\hat{x}}=x\, \vec{e_{x}}+y\, \vec{e_{y}}+z\, \vec{e_{z}},
\end{equation}
then the wave-center of the spheron can be recovered with
\begin{equation}
\vec{o}=\int\psi^{*} \, \vec{\hat{x}} \,\psi \, dV,
\end{equation}
a quantity depending on the frequency is delivered by 
\begin{eqnarray}
\frac{\omega}{\gamma_s} & = &
\int\psi^{*}\left(-i\frac{\partial}{\partial t}\right)\psi \, dV
\end{eqnarray}
and a quantity depending on the velocity is obtained with
\begin{equation}
\frac{\omega}{\gamma_s}\frac{\mathrm{1}}{c_s^{2}}\vec{v}=\int\psi^{*} \, i \, \vec{\nabla}
\, \psi \,dV.
\end{equation}
These operators are close to the quantum-mechanical operators
for location, energy and momentum respectively.

Other analogies can be found, but discussing them here will not add any
more insight concerning the subject of this contribution. Nevertheless
one last remark is in order: since the wave equation results from
applying classical mechanics to the medium, it is perfectly possible
to employ Hamiltonians or Lagrangians, as is common practice in
quantum mechanics.

\subsection{A spheronic theory of relativity}
\label{sec:SpherRelat}

The Lorentz composition results in relativistic effects for the
amplitude and quantum mechanical effects for the wave part at the same
time. The spheron can only be fully understood by considering both
aspects at the same time, it is a phase wave and a ``localized''
amplitude represented by a point-like wave center, one entity with a
``dualistic'' nature. This concludes our reasoning for choosing
spherons as representing matter with rest mass in our toy universe.

Contraction and dilation provide a direct path to special relativity,
including the 4-dimensional space-time, and the full apparatus of
relativity. To get there, it is necessary to give up the human
perspective and adopt the perspective of a spheronic observer.

First we note that a spheronic observer has no means to measure her/his
state of motion against the gas. If he would construct a "spheronic
interferometer" to measure her/his speed against a hypothesized "ether",
he would end up with a null-result, as did Michelson and Morley with
their experiments to measure an ether drift in the late 19th
century. This can be shown by an explicit calculation, but it is
easier to remind oneself that the only reason to introduce the
contraction of rods and the dilation of clocks were the need to
explain these unexpected null-results.

With the ideas of rigid bodies in mind, these indeed seem to be
awkward ad-hoc explanations with little physical plausibility. In his
1908 paper, Minkowski acknowledges them to work mathematically but
calls them a ``present from above'' (``Geschenk von oben'', Minkowski
1908). He then states, that Lorentz' idea is completely equivalent
("v\"ollig \"aquivalent") to his new conception of time and space.

From the perspective of a human physicist, an analogue to a Lorentzian
ether theory is the best choice to describe the vibrations of the
medium. Note, that the relativistic contraction and dilation in the idealized
universe are no ``present from above'', but a direct consequence of
the wave nature (``Wellenartigkeit'') of matter.

From the perspective of an internal spheronic observer, things are
different. What is an obvious and measurable contraction of the
spheronic observer's standard ruler for an external human ({\it who is
  not an observer made of spheronic matter and who therefore can
  observe the idealized universe from ``outside'' by not being part of
  it}) is actually unmeasurable and unnoticeable to the spheronic
observer.\footnote{Considering Figure 2 it might be thought that an
  observer may be able to determine her/his state of motion by
  measuring the distortion of her/his spheronic waves, because a
  spheron appears contracted in the direction of motion. Thus, taking
  the standard rod to be the distance between the centre and the first
  wave maximum of a spheron, it follows that this unit length appears
  longer perpendicularly to the direction of motion. This, however, is
  only the case for an external observer who is not made of spherons
  and is not part of the spheronic universe. When the spheronic
  observer rotates the rod by 90 degrees into the direction of motion,
  the rod contracts just the same as the spheronic observer does,
  since the rod is made equally of spherons. Thus, an observer made of
  spheronic matter cannot measure the length contraction with her/his
  tools which are made of spheronic matter and therefore the motion
  relative to the medium cannot be determined (see also Appendix C).}
The same applies to the dilation of proper time. The spheronic
observer's standard ruler and clock define her/his units of
measurement. Putting oneself in the position of such an observer with
the task of developing a physical theory leads to a spheronic theory
of relativity.

A human observer in the real universe thinking up the idealized universe can
easily switch between these two perspectives. This makes it possible
to explore all the seemingly paradoxical results of relativity within
the classical framework of our idealized universe. This includes, among
other issues, the relativity of simultaneity and the twin
paradox. With a set of comoving synchronized spheronical co-observers
it is also possible to setup a 4-dimensional spheronic space-time on
that basis.

Beyond that it might be interesting to note that the intimate relation
between time and space measured and experienced by a spheronic
observer is deeply rooted in the material (i.e. spheronic) nature of
her/his existence.  It is for this reason, that a spheronic observer
always measures the speed of sound to be a constant quantity.

But the properties of the matter the observer is made of determine the
reality of a spheronic observer to an even greater extent. The
material layer effectively shields a spheronic observer from the
underlying physical reality and cannot be circumvented by him. As a
result e.g. the existence of the full fledged wave function is, for a
spheronic observer, not measurable and remains a non-physical (sub-
trans- or meta-physical) speculation for him. The reach of her/his
material tools define the reach of her/his physical theories.

\section{DISCUSSION}
\label{sec:disc}

The gedanken experiment presented here, in which an idealized universe
is created from simple classical principles, indicates intriguing
properties that resemble some important aspects of our real
world. These aspects may be relevant for a quantum mechanical and
special relativistic understanding of our real universe.

In our real universe, we are in a position which is in many respects
similar to the situation of spheronic observers in the idealized
universe. Like them we are made of matter which can be described as
waves, and such are all our tools and means to gather information about
the universe we inhabit. 

Turning back to our idealized world, we have thus discovered that the very
simple approach of analyzing a known solution of the wave equation
with tools usually not applied there can reveal many elements of
modern physics. Among these are a natural speed limit for entities,
relativistic contraction and time dilation as well as supersonic
(superluminal) phase waves and an example how a ``ridiculous looking
proposal'' (Penrose 2005, page 500) of an operator-logic can be
founded in geometrical properties of the object under study.

The reader will have noticed a major difference between a "de Broglie
wave" associated to a spheron and the de Broglie wave associated with
a particle in traditional quantum mechanics. Following a paradigm
already present in de Broglie's nobel-prize winning thesis,
traditional quantum mechanical models treat a particle as a
wave-packet to arrive at a localized entity. This packet is made of a
continuous spectrum of de Broglie waves and although there cannot be
any experimental evidence for this it still seems to be the standard
model (e.g. Tipler \& Llewellyn 2002). Within this scheme, quantum
mechanics needs to postulate for every class of elementary particle a
separate field. In contrast to this, a spheron is localized by its
amplitude, which is "associated" to the "de Broglie wave" by
multiplication. As a consequence of this, our idealized quantum
mechanics needs just one "field" with just one governing "universal"
equation and the various types of particles are modelled as different
solutions to this. In the case of the spheron, the amplitude
complements the "de Broglie wave" and both together solve the wave
equation, while the "de Broglie wave" only solves the
"Klein-Gordon-equation".

Naturally, we emphasized more the similarities than the differences
between spherons and matter, of which there are plenty. To mention
only one: Spherons do not interact with any other vibrations, be it
another spheron or a plane wave. This is due to the linearity of the
wave equation which leads to the possibility of superimposing
solutions to generate a new one. Other properties of real matter are
missing completely. In the real universe, there is definitely more
than one type of elementary particle and there are properties like
spin, which have no counterpart in spherons.

Some of these differences might be studied within the framework of the
idealized universe by varying the properties of the medium, since this
generally determines the existence, the properties of and the
interaction between the wave solutions that can exist within it.
Giving up on the linearity of the wave equation means giving up the
superposition principle and would thus enforce interacting
entities. Keeping the linearity but dropping the requirement of a
curl-free medium allows for it to harbor more diverse classes of
vibrations including phenomena with spin-like properties.

It is generally an interesting endeavor to hunt for analogies and
various more or less obvious possibilities to do so provide a vast
playground. Whatever the odds are to develop a model that comes closer
to observed reality: it seems remarkable that even such a most simple
medium as an idealized gas can reproduce so much of the fundamental
elements of modern physics.

Our gedanken experiment may thus shed an interesting light on
interpreting the quantum mechanical and special relativistic
properties of the real universe, in the sense that these two
properties may be two sides of the same coin. But, perhaps more
importantly, the present spheron-gedanken experiment may be useful for
teaching quantum mechanics and special relativity, as a means of
visualizing the possibly intimate connection between the wave-like
properties of matter and special relativity, in the sense that if real
matter can be described by waves then special relativity automatically
emerges.

\begin{acknowledgements}
  We thank the referee for very careful analysis of the material
  presented here and for helpful suggestions which clarified the
  arguments. The challenge discussed in Appendix~C has been posed by
  the referee.
\end{acknowledgements}

\begin{appendix}

\section{APPENDIX A - REAL DE-BROGLIE WAVES}

To make the relationship between the wave part of a moving spheron in
our idealized universe and a quantum-mechanical de Broglie wave in the real
universe (in which the limiting speed is the speed of light, $c$)
explicit, note that a de Broglie wave in the real universe associated
with a particle of rest mass $m_{0}$, moving in the direction of
increasing values of $z$, is usually written as
\begin{equation}
B\left(x,t\right)=e^{i\left(pz-Et\right)/\hbar},\label{eq:de-Broglie-Welle}
\end{equation}
where $E=\hbar w$ is the relativistic energy and $p=\hbar k$
is the relativistic momentum. The energy is 
\begin{eqnarray}
E & = & mc^{2}=\hbar w,\\
 & = & \frac{m_{0}}{\sqrt{1-\frac{v^{2}}{c^{2}}}}c^{2}=\hbar\frac{w_{0}}{\sqrt{1-\frac{v^{2}}{c^{2}}}},
\end{eqnarray}
and the momentum is
\begin{eqnarray}
p & = & mv=\frac{E}{c^{2}}v=\hbar k,\\
 & = & \frac{m_{0}}{\sqrt{1-\frac{v^{2}}{c^{2}}}}v=\hbar\frac{w_{0}}{\sqrt{1-\frac{v^{2}}{c^{2}}}}\frac{1}{c^{2}}v=\hbar k.
\end{eqnarray}
Hence
\begin{eqnarray}
Et-pz & = & \hbar\frac{w_{0}}{\sqrt{1-\frac{v^{2}}{c^{2}}}}t-\hbar\frac{v}{c^{2}}\frac{w_{0}}{\sqrt{1-\frac{v^{2}}{c^{2}}}}z,\\
 & = & \hbar\frac{w_{0}}{\sqrt{1-\frac{v^{2}}{c^{2}}}}\left(t-\frac{v}{c^{2}}z\right).
\end{eqnarray}
Thus, from (\ref{eq:de-Broglie-Welle}), 
\begin{eqnarray}
B\left(x,t\right) & = & e^{-i\left(\frac{w_{0}}{\gamma_s}\left(t-\frac{v}{c^{2}}z\right)\right)},\\
 & = & e^{-i\left(w_{0}t\mbox{'}\right)},\\
 & = & {\rm cos}\left(-w_{0}t\mbox{'}\right)+i\, {\rm sin}\left(-w_{0}t\mbox{'}\right),
\end{eqnarray}
for a de Broglie wave. Compare this to 
\begin{equation}
B\left(t'\right)={\rm cos}\left(\omega t'\right),\label{eq:reale-de-Broglie-welle}
\end{equation}
(see eqs~\ref{eq:mov-spheron-expl} and~\ref{eq:mov-spheron}),
which is the plane wave part of the moving spheron. Since ${\rm
  cos}\left(-\alpha\right)={\rm cos}\left(\alpha\right)$
(\ref{eq:reale-de-Broglie-welle}) may thus be seen as the real part of
a de Broglie type plane wave.

\section*{APPENDIX B - VOLUME INTEGRALS AND OPERATORS}

Based on (eqs~\ref{eq:gamma}-\ref{eq:rs}) and assuming there is a 
speed limit $c$ (which is $c_s$ in our idealized universe or the speed of light in
the real universe)
\begin{eqnarray}
r' & = & \left(x^{2}+y^{2}+z^{'^{2}}\right)^{\frac{1}{2}},\\
 & = & r\left(1+\frac{z^{'^{2}}-z^{2}}{r^{2}}\right)^{\frac{1}{2}},\\
 & = & r\left(1+\frac{\frac{\left(z-vt\right)^{2}}{1-\frac{v^{2}}{c^{2}}}-z^{2}}{r^{2}}\right)^{\frac{1}{2}},
\end{eqnarray}
can be concluded, and at $t=0$ 
\begin{eqnarray}
r' & = & r\left(1+\frac{\frac{z^{2}}{1-\frac{v^{2}}{c^{2}}}-z^{2}}{r^{2}}\right)^{\frac{1}{2}},\\
 & = & r\left(1+\frac{\frac{1-\left(1-\frac{v^{2}}{c^{2}}\right)}{1-\frac{v^{2}}{c^{2}}}z^{2}}{r^{2}}\right)^{\frac{1}{2}},\\
% & = & r\left(1+\left(\frac{\frac{v^{2}}{c^{2}}}{1-\frac{v^{2}}{c^{2}}}\right)\frac{z^{2}}{r^{2}}\right)^{\frac{1}{2}},
\end{eqnarray}
holds. In spherical coordinates with 
\begin{equation}
{\rm cos}\left(\theta\right)=\frac{z}{r},
\end{equation}
this may be written as 
\begin{equation}
r'=r\left(1+\left(\frac{1-\gamma_s^{2}}{\gamma_s^{2}}\right){\rm cos}^{2}\left(\theta\right)\right)^{\frac{1}{2}}.
\end{equation}
Therefore
\begin{equation}
dr'=dr\left(1+\left(\frac{1-\gamma_s^{2}}{\gamma_s^{2}}\right){\rm cos}^{2}\left(\theta\right)\right)^{\frac{1}{2}},
\end{equation}
or
\begin{eqnarray}
r & = & \frac{r'}{\left(1+\left(\frac{1-\gamma_s^{2}}{\gamma_s^{2}}\right){\rm cos}^{2}\left(\theta\right)\right)^{\frac{1}{2}}},\\
dr & = & \frac{dr'}{\left(1+\left(\frac{1-\gamma_s^{2}}{\gamma_s^{2}}\right){\rm cos}^{2}\left(\theta\right)\right)^{\frac{1}{2}}},
\end{eqnarray}
holds. With $\rho=2A_{0}\frac{{\rm sin}\left(kr'\right)}{r'}e^{i\omega t'}$,
the volume integral over $\rho\rho^{*}$ is
\begin{eqnarray}
\int\rho\rho^{*}dV & = & \int\int\int A^{2}\left(r'\right)r^{2}{\rm sin}\left(\theta\right)drd\theta d\phi,\\
 & = & \int\int\int
 A^{2}\left(r'\right)\frac{r^{'^{2}}}{\left(1+\left(\frac{1-\gamma_s^{2}}{\gamma_s^{2}}\right){\rm
       cos}^{2}\left(\theta\right)\right)^{\frac{3}{2}}} \nonumber \\
& \times & {\rm sin}\left(\theta\right)dr'd\theta d\phi.
\end{eqnarray}
Because
\begin{equation}
\int_{0}^{\pi}\frac{{\rm
    sin}\left(\theta\right)}{\left(1+\left(\frac{1-\gamma_s^{2}}{\gamma_s^{2}}\right){\rm
      cos}^{2}\left(\theta\right)\right)^{\frac{3}{2}}}d\theta
=\frac{2}{\left(1+\frac{1-\gamma_s^{2}}{\gamma_s^{2}}\right)^{\frac{1}{2}}}=2\gamma_s,
\end{equation}
and 
\begin{equation}
\int_{0}^{2\pi}d\phi=2\pi,
\end{equation}
the volume integral is
\begin{eqnarray}
\int\rho\rho^{*}dV & = & 2\gamma_s 2\pi\int A^{2}\left(r'\right)r^{'^{2}}dr',\\
 & = & 2\gamma_s 2\pi\int\left(2A_{0}\frac{{\rm sin}\left(kr'\right)}{r'}\right)^{2}r^{'^{2}}dr',\\
 & = & 2\gamma_s 2\pi4A_{0}^{2}\int {\rm sin}^{2}\left(kr'\right)dr',\\
 & = & 2\gamma_s 2\pi4A_{0}^{2}\left[\frac{1}{2}r'-\frac{1}{4k}{\rm sin}\left(2kr'\right)\right]_{0}^{R'},
\end{eqnarray}
where the last step follows from integrating from zero to $R'$. Define
\begin{equation}
\vartheta'=\frac{1}{2A_{0}}\frac{1}{\pi r'}\sqrt{\frac{1}{2k\gamma_s}},
\end{equation}
 set 
\begin{eqnarray}
\chi & =\vartheta' & 2A_{0}\frac{{\rm sin}\left(kr'\right)}{r'} =
\vartheta' \, A(r'),\\
 & = & \frac{1}{\pi r'}\sqrt{\frac{1}{2k\gamma_s}}\frac{{\rm sin}\left(kr'\right)}{r'},
\end{eqnarray}
and 
\[
\psi=\chi e^{i\omega t},
\]
\noindent then 
\begin{equation}
\int\psi^{*}\psi dV=\int\chi^{2}dV.
\end{equation}
The volume integral over $\psi^{*}\psi$ is
\begin{eqnarray}
\int\chi^{2}dV & = & \int\int\int\left(\frac{1}{\pi
    r'}\sqrt{\frac{1}{2k\gamma_s}}\frac{{\rm
      sin}\left(kr'\right)}{r'}\right)^{2}\\
& \times &r^{2}{\rm sin}\left(\theta\right)drd\theta d\phi,\\
 & = &
 \frac{1}{\pi^{2}2k\gamma_s}\int\int\int\left(\frac{1}{r'}\frac{{\rm
       sin}\left(kr'\right)}{r'}\right)^{2}\\
& \times &r^{2}{\rm sin}\left(\theta\right)drd\theta d\phi,\\
 & = & \frac{2\gamma_s 2\pi}{\pi^{2}2k\gamma_s}\int\left(\frac{1}{r'}\frac{{\rm sin}\left(kr'\right)}{r'}\right)^{2}r^{'^{2}}dr',\\
 & = & \frac{2}{\pi k}\int\left(\frac{{\rm sin}\left(kr'\right)}{r'}\right)^{2}dr',\\
 & = & \frac{2}{k\pi}\left(-\left[\frac{{\rm
         sin}^{2}\left(kr'\right)}{r'}\right]_{0}^{R'} \right. \\
& + & \left. k\int_{0}^{R'}\frac{{\rm sin}\left(2kr'\right)}{r'}dr'\right).
\end{eqnarray}
The integral 
\begin{equation}
\int_{r'}\left(\frac{{\rm sin}\left(kr'\right)}{r'}\right)^{2}dr',
\end{equation}
thus approaches, for sufficiently large $R'$ and to arbitrary
precision, $k\frac{\pi}{2}$ and the integral of $\psi^{*}\psi$ then
comes out as
\begin{equation}
\int\psi^{*}\psi \, dV=1.
\end{equation}
At any other time $t\neq0$, the calculations hold mutatis mutandis,
since the values of $\chi^{2}$ are displaced by an amount of $z=vt$ on
the z-axis, which is equivalent to a shift of origin, and which does
not alter the summation over all of space.  With such a ``normalized''
wave like $\psi$, the effect of the operator
\begin{equation}
\vec{\hat{x}}=x\vec{e_{x}}+y\vec{e_{y}}+z\vec{e_{z}},
\end{equation}
in 
\begin{eqnarray}
\vec{o} & = & \int\psi^{*} \, \vec{\hat{x}} \, \psi \, dV,\\
 & = & \int\vec{\hat{x}} \, \chi^{2} \, dV,
\end{eqnarray}
is the same as the calculation of an average of a statistically
distributed quantity and returns the wave center as an ``average'' of
position.  With
\begin{eqnarray}
e^{-i\omega t'}i\vec{\nabla}e^{i\omega t'} & = & e^{-i\omega t'}e^{i\omega t'}i\vec{\nabla}\left(i\omega t'\right),\\
 & = & \frac{\omega}{\gamma_s}\frac{\mathrm{1}}{c^{2}}\vec{v},
\end{eqnarray}
the volume integral of $\psi^{*}i\vec{\nabla}\psi$ becomes
\begin{eqnarray}
\int\psi^{*} \, i \, \vec{\nabla} \, \psi \, dV & = & \int\chi e^{-i\omega t'}
\nonumber \\
&\times &i\, \left(\vec{\nabla}\chi e^{i\omega t'}+\chi\vec{\nabla}e^{i\omega t'}\right)dV,\\
 & = & i\int\chi\vec{\nabla}\chi dV \nonumber \\
& + &i\int\chi^{2}e^{-i\omega t'}\vec{\nabla}e^{i\omega t'}dV,\\
 & = & \frac{i}{2}\int\vec{\nabla}\chi^{2}dV \nonumber \\
& + &\frac{\omega}{\gamma_s}\frac{\mathrm{1}}{c^{2}}\vec{v}\int\chi^{2}dV,\\
 & = & \frac{\omega}{\gamma_s}\frac{\mathrm{1}}{c^{2}}\vec{v},
\end{eqnarray}
where 
\begin{equation}
\int\vec{\nabla}\chi^{2}dV=0
\end{equation}
has been used. This follows here from the following considerations:
The integrand is the gradient of a scalar quantity. Using Gauss' theorem,
such a volume integral may be written as a surface integral, in general
\begin{equation}
\int\nabla\Phi\left(\vec{x}\right)dV=\oint\Phi\left(\vec{x}\right)d\vec{A}
\end{equation}
holds, where $\Phi\left(\vec{x}\right)$ is an arbitrary
scalar field and $d\vec{A}$ is the surface normal of the (closed)
surface enclosing the volume.

Integrating in three dimensions over a spherical volume centered at
the origin, $d\vec{A}$ is parallel to the position vector
$\vec{x}$, hence 
\begin{equation}
d\vec{A}\left(-\vec{x}\right)=-d\vec{A}\left(\vec{x}\right).
\end{equation}
For the scalar field $\chi^{2}\left(\vec{x}\right)$
at $t=0$,
\begin{equation}
\chi^{2}\left(\vec{x}\right)=\chi^{2}\left(-\vec{x}\right)
\end{equation}
holds, from which 
\begin{equation}
-\chi^{2}\left(\vec{x}\right)d\vec{A}\left(\vec{x}\right)=\chi^{2}\left(-\vec{x}\right)d\vec{A}\left(-\vec{x}\right)
\end{equation}
can be concluded. The surface integral of $\chi^{2}$ is
therefore
\begin{equation}
\oint_{\partial A}\chi^{2}d\vec{A}=0,
\end{equation}
from which 
\begin{equation}
\int\nabla\chi^{2}dV=0
\end{equation}
follows. For any time $t\neq0$, this calculation holds
mutatis mutandis. With 
\begin{eqnarray}
e^{-i\omega t'}\left(-i\frac{\partial}{\partial t}\right)e^{i\omega  t'} 
& = & e^{-i\omega t'}e^{i\omega t'} \nonumber \\
&\times& \left(-i\frac{\partial} {\partial t}\right)\left(i\omega t'\right),\\
 & = & \frac{\omega}{\gamma_s},
\end{eqnarray}
it follows that
\begin{eqnarray}
\int\psi^{*}\left(-i\frac{\partial}{\partial t}\right)\psi dV & = &
-\int\chi e^{-i\omega t'}i \nonumber \\
&& \left(\frac{\partial}{\partial t}\chi e^{i\omega
    t'}+\chi\frac{\partial}{\partial t}e^{i\omega t'}\right)\,dV, \nonumber\\ 
& & \\
 & = & -i\int\chi\frac{\partial}{\partial t}\chi dV \nonumber \\
&-&i\int\chi^{2}e^{-i\omega t'}\frac{\partial}{\partial t}e^{i\omega
  t'}\, dV, \nonumber \\ 
& &\\
 & = & -\frac{i}{2}\int\frac{\partial}{\partial t}\chi^{2}dV \nonumber
 \\
&+&\frac{\omega}{\gamma_s}\int\chi^{2}dV,\\
 & = & \frac{\omega}{\gamma_s},
\end{eqnarray}
where 
\begin{equation}
\int\frac{\partial}{\partial t}\chi^{2}dV=0
\end{equation}
has been used. Using Leibniz' integral rule
\begin{equation}
\frac{\partial}{\partial x}\intop_{y_{0}}^{y_{1}}f(x,y)dy=\intop_{y_{0}}^{y_{1}}\frac{\partial}{\partial x}f(x,y)dy,
\end{equation}
it can be concluded, that
\begin{equation}
\int\frac{\partial}{\partial t}\chi^{2}dV=\frac{\partial}{\partial t}\int\chi^{2}dV.
\end{equation}
Now 
\begin{eqnarray}
\frac{\partial}{\partial t}\int\chi^{2}dV & = & \frac{\partial}{\partial t}1,\\
 & = & 0
\end{eqnarray}
which may be regarded as a conservation law for the spheron.

\section*{APPENDIX C - A THOUGHT EXPERIMENT}

Here a gedanken experiment is discussed which may rightfully be raised
in an attempt to disprove that the classical spheronic universe may
have special relativistic properties.

Let us propose a thought experiment to show that, contrary to the
assertion made in Section~\ref{sec:SpherRelat}, a spheronic observer
can measure her/his or her state of motion against the gas. As the
calculations below demonstrate, this is not possible.

\underbar{Challenge}: Assume that spheronic observers can perform
radar-ranging: they can emit a wave travelling at the speed of sound
and measure the time until its return (this is, so to speak, a
``spheronic photon'': the plane wave as discussed in
Sec.~\ref{sec:Lor}). Now, in a rest frame (not necessarily the gas
rest frame), an external human observer (i.e. we) places A and C a
certain distance apart, and B is put in the middle. We can construct
this scenario by, for example, telling B to stay still and having A
and C move away until a radar range of (for example) 1 unit of
eigen-time is observed. A, B and C are stationary with respect to each
other. Again, the external human observer can test this by seeing if
repeated radar-rangings have a constant return time.

Now, the external human observer gives A and C the following
instructions: according to time on her/his  wristwatch,
``spheronic photons'' (i.e. plane parallel waves in the idealized
  spheronic universe) are emitted towards B at the rate of one per unit of
eigen-time. If A, B and C were in the gas rest frame, then B would
observe that A and C’s spheronic photons arrive at the same rate. But, if their
rest frame is moving with respect to the gas rest frame, then {\it
  spheronic photons travelling down wind are expected to arrive more frequently}.

By symmetry, the distances AB and BC are the same: this holds true of
all radar ranging, even if the outward and inward journey are at
different speeds. By symmetry, the internal spheronic observers know
that A, B and C’s watches tick at the same rate: they are all in the
same rest frame i.e. they all have the same velocity with respect to
the gas. Thus, they must conclude that ``spheronic photons'' travel
faster in certain directions. They can measure their motion against
the gas.

This is not the case in special relativity of our (human) real
world. In any frame, if A, B and C are set up by radar ranging with
light, and then have them send B light pulses at a rate of one per
second, then B will always see the same pulse frequency from A and C.

Thus, the speed of sound appears to not be invariant for spheronic
observers. There appears to be no spheronic theory of relativity.
Another way of seeing this is in the swap from ``Lorentz
transformation'' to ``Lorentz composition of quantities''
(eqs~\ref{eq:gamma} to~\ref{eq:rs}). If quantities are only composed,
then the final results don't necessarily have physical meaning. One
would be putting together variables in a mathematically interesting
way. There is no rationale for interpreting the primed coordinates as
anything, much less the observed space and time of a moving spheronic
observer.

\underbar{Answer}: Spheronic observers limited to and made of the
``matter'' (i.e. spherons) of the idealized universe cannot detect a
difference in the speed of sound because the frequency shift in either
direction is always accompanied by a corresponding change in measured
length, given that the measurements need to be made with a standard
ruler made of spheronic matter.

In particular, the above statement that ``{\it spheronic photons
  travelling downwind will arrive more frequently.}'' is not
correct. Both sets of spheronic photons will arrive at the same rate
they are emitted, namely at one per spheronic unit time. What is
correct is the following: spherons travelling downwind will arrive
with a higher relative velocity, but only as seen by the external
(human) observer.  {\it This is the point where the intuition is
  misleading since we are not used to adopting the point of view of a
  spheronic observer. This observer is restricted to her/his spheronic
  tools}.  And this raises the question of how B will measure
velocities. This obviously touches on time dilation and length
contraction.

This is clarified with an explicit calculation: Let us begin
with the case of an internal (spheronic) observer (B) at rest with
respect to the gas. Without loss of generality, let her/his
spheron defining the observer's clock and rulers be described by
$\frac{sin\left(kr\right)}{r}cos\left(\omega t\right)$, let her/his
wrist watch be based on the oscillations in the wave center,
i.e. $cos\left(\omega t\right)$, and let her/his rulers defining unity
stretch from the wavecenter to the first zero of the spheron. Let a
minimal time unit of the spheronic observer be the time elapsed
between two extrema of $cos\left(\omega t\right)$, measuring time than
means for the internal spheronic observer counting the ticks of
the wrist watch. The time between two subsequent ticks as measured by
the external human scientist is then $\pi/\omega$, the length of her/his
or her ruler as measured by the external human scientist is
$\pi/k$. These units of space and time may be transformed into each
other with the help of the speed of sound. The relation

\[
\frac{\pi}{k}=c_{s}\frac{\pi}{\mbox{\ensuremath{\omega}}}
\]

\noindent always holds, since $\omega^{2}/k^{2}=c_{s}^{2}$ follows from the
spheron having to obey the wave-equation. So placing two internal
fellow-spheronic-observers (A and C) at the respective ends of two
rulers stretching in opposite directions is equivalent to placing them
one ``tick'' away as a result of a radar measurement. Now let these
three spheronic observers synchronize their clocks the Einstein way
and provide instructions to the two outer observers to send a sound
wave in the direction of the spheronic observer in the middle at a
distinctive reading $\tau_{d}$ of their wrist watches and repeat that
process after one tick on their respective wrist watches. Again
without loss of generality, let $\tau_{d}=0$, the spheronic
observer in the middle will register at proper time $\tau_{d}=1$
incoming waves from both sides. For the external human scientist the
waves will reach the middle observer at a delay of $\Delta
t=\pi/\omega$ after they have been started. These calculations hold
true for all subsequently sent waves at each tick of proper time.

Now let's make things moving and use a propagating spheron as the
basis of the clock and rulers of a spheronic observer. Using the
``Lorentz-composition'' of quantities (Sec.~\ref{sec:Lor})

\begin{eqnarray*}
\gamma_{s} & = & \sqrt{1-\frac{v^{2}}{c_{s}^{2}}},\\
t^{'} & = & \frac{1}{\gamma_{s}}\left(t-\frac{v}{c_{s}^{2}}z\right),\\
z^{'} & = & \frac{1}{\gamma_{s}}\left(z-vt\right),\\
r^{'} & = & \left(x^{2}+y^{2}+z^{'^{2}}\right)^{\frac{1}{2}},
\end{eqnarray*}

\noindent then

\[
\rho = \frac{sin\left(kr^{'}\right)}{r^{'}}cos\left(\omega t^{'}\right)
\]

\noindent is a solution of the wave equation. As shown, the wave center moves
with speed $v$ in the direction of increasing $z$, the oscillation
in the wave center is then

\begin{eqnarray*}
cos\left(\omega t^{'}\right) & = & cos\left(\omega\frac{1}{\gamma_{s}}\left(t-\frac{v}{c_{s}^{2}}vt\right)\right),\\
 & = & cos\left(\omega\frac{t}{\gamma_{s}}\left(1-\frac{v^{2}}{c_{s}^{2}}\right)\right),\\
 & = & cos\left(\omega\gamma_{s}t\right)
\end{eqnarray*}

The external human scientist will then measure the time between two
ticks of the spheronic observer's wrist watch as
$\pi/\left(\gamma_{s}\omega\right)=\pi/\left(\sqrt{1-\frac{v^{2}}{c_{s}^{2}}}\omega\right)$
which is longer than the time between two ticks of the observer at
rest and is equivalent to saying that the rate of ticks has gone down
which constitutes a dilation of proper time. Constructing the rulers
in the same way as the case of the spheronic observer at rest leads to
a contraction of the rulers which are parallel to the direction of
movement. This can be seen by setting $kr^{'}=\pi$ and solving for
$z$. With $x=y=0$ this leads to $z=\gamma_{s}\pi/k$. The spheronic
observer performing radar measurements of her/his rulers will find
that all her/his rulers have a length of one tick of her/his proper
time, as can be shown by an explicit calculation. Now let us place two
comoving spheronic observers at the end of the respective rulers, have
them synchronize their watches by the same procedure and provide them
with the same set of instructions.  The Einstein synchronization leads
to the wrist watches showing the above $t^{'}$ as proper time $\tau$,
the instructions given are now sending a sound wave in the direction
of the middle observer at $\tau=0$ and then subsequently at each tick
of proper time. Without loss of generality, let the middle spheronic
observer be in the center of the external human scientist's coordinate
system at $t=0$ and let us assume that her/his wrist watch then shows
$\tau=0$. The position of the observer with a lower z-value (the
``lower'' observer) in general is given by
$x=-\gamma_{s}\frac{\pi}{k}+vt$ , so her/his or her wrist watch will
then show a proper time of

\[
t^{'}=\frac{1}{\gamma_{s}}\left(t-\frac{v}{c_{s}^{2}}\left(-\gamma_{s}\frac{\pi}{k}+vt\right)\right).
\]

\noindent  To find the time where the lower observer emits the sound wave, the
``lower'' proper time has to be set to zero:

\begin{eqnarray*}
0 & = & \frac{1}{\gamma_{s}}\left(t-\frac{v}{c_{s}^{2}}\left(-\gamma_{s}\frac{\pi}{k}+vt\right)\right),\\
 & = & \frac{1}{\gamma_{s}}\left(\frac{v}{c_{s}^{2}}\gamma_{s}\frac{\pi}{k}+t\left(1-\frac{v^{2}}{c_{s}^{2}}\right)\right),\\
-\gamma_{s}t & = & \frac{v}{c_{s}^{2}}\frac{\pi}{k},\\
t & = & -\frac{1}{\gamma_{s}}\frac{v}{c_{s}^{2}}\frac{\pi}{k}
\end{eqnarray*}
at this time her/his position was 

\begin{eqnarray*}
z & = & -\gamma_{s}\frac{\pi}{k}-v\frac{1}{\gamma_{s}}\frac{v}{c_{s}^{2}}\frac{\pi}{k},\\
 & = & -\gamma_{s}\frac{\pi}{k}\left(1+\frac{v^{2}}{c_{s}^{2}}\frac{1}{\gamma_{s}^{2}}\right),\\
 & = & -\gamma_{s}\frac{\pi}{k}\left(\frac{\gamma_{s}^{2}+\frac{v^{2}}{c_{s}^{2}}}{\gamma_{s}^{2}}\right),\\
 & = & -\gamma_{s}\frac{\pi}{k}\left(\frac{1}{\gamma_{s}^{2}}\right),\\
 & = & -\frac{\pi}{k}\frac{1}{\gamma_{s}}.
\end{eqnarray*}
Calculating in the same way the parameter for the upper observer
leads to the emission of the sound wave at 

\begin{eqnarray*}
t & = & \frac{1}{\gamma_{s}}\frac{v}{c_{s}^{2}}\frac{\pi}{k},\\
z & = & \frac{\pi}{k}\frac{1}{\gamma_{s}}.
\end{eqnarray*}
To find the travelling time of the lower sound wave 

\[
-\frac{\pi}{k}\frac{1}{\gamma_{s}}+c_{s}\Delta t=-\frac{\pi}{k}\frac{1}{\gamma_{s}}+\gamma_{s}\frac{\pi}{k}+v\Delta t
\]
 has to be solved for $\Delta t$ which leads to 

\[
\Delta t=\frac{\gamma_{s}\frac{\pi}{k}}{c_{s}-v}.
\]
The sound wave will then meet the middle observer at 

\begin{eqnarray*}
z & = & -\frac{\pi}{k}\frac{1}{\gamma_{s}}+c_{s}\frac{\gamma_{s}\frac{\pi}{k}}{c_{s}-v},\\
 & = & \gamma_{s}\frac{\pi}{k}\left(\frac{1}{1-\frac{v}{c_{s}}}-\frac{1}{\gamma_{s}^{2}}\right),\\
 & = & \gamma_{s}\frac{\pi}{k}\left(\frac{1}{1-\frac{v}{c_{s}}}-\frac{1}{\left(1-\frac{v}{c_{s}}\right)\left(1+\frac{v}{c_{s}}\right)}\right),\\
 & = & \gamma_{s}\frac{\pi}{k}\left(\frac{\frac{v}{c_{s}}}{\gamma_{s}^{2}}\right),\\
 & = & \frac{\pi}{k}\left(\frac{\frac{v}{c_{s}}}{\gamma_{s}}\right).
\end{eqnarray*}

The travelling time for the upper sound wave is derived from 
\[
\frac{\pi}{k}\frac{1}{\gamma_{s}}-c_{s}\Delta t=\frac{\pi}{k}\frac{1}{\gamma_{s}}-\gamma_{s}\frac{\pi}{k}+v\Delta t
\]
 and leads to 
\[
\Delta t=\frac{\gamma_{s}\frac{\pi}{k}}{c_{s}+v},
\]
so the upper wave meets the middle observer at 
\begin{eqnarray*}
z & = & \frac{\pi}{k}\frac{1}{\gamma_{s}}-c_{s}\frac{\gamma_{s}\frac{\pi}{k}}{c_{s}+v},\\
 & = & \gamma_{s}\frac{\pi}{k}\left(-\frac{1}{1+\frac{v}{c_{s}}}+\frac{1}{\gamma_{s}^{2}}\right),\\
 & = & \gamma_{s}\frac{\pi}{k}\left(-\frac{1}{1+\frac{v}{c_{s}}}+\frac{1}{\left(1-\frac{v}{c_{s}}\right)\left(1+\frac{v}{c_{s}}\right)}\right),\\
 & = & \gamma_{s}\frac{\pi}{k}\left(\frac{\frac{v}{c_{s}}}{\gamma_{s}^{2}}\right),\\
 & = & \frac{\pi}{k}\left(\frac{\frac{v}{c_{s}}}{\gamma_{s}}\right).
\end{eqnarray*}
Thus both soundwaves are registered at the same place. The external human scientist
will record this happening at 
$t=\frac{\pi}{k}\left(\frac{\frac{1}{c_{s}}}{\gamma_{s}}\right)=\frac{\pi}{\omega\gamma_{s}}$,
which is the time he or she measured before for one tick of the spheronic
observer's proper time. Indeed, the middle observer's wrist watch
then shows a proper time of 
\begin{eqnarray*}
\tau & = & \frac{1}{\gamma_{s}}\left(\frac{\pi}{k}\left(\frac{\frac{1}{c_{s}}}{\gamma_{s}}\right)-\frac{v}{c_{s}^{2}}\frac{\pi}{k}\left(\frac{\frac{v}{c_{s}}}{\gamma_{s}}\right)\right),\\
 & = & \frac{1}{\gamma_{s}}\frac{\pi}{k}\left(\frac{\frac{1}{c_{s}}}{\gamma_{s}}\right)\left(1-\frac{v^{2}}{c_{s}^{2}}\right),\\
 & = & \frac{\pi}{k}\frac{1}{c_{s}},\\
 & = & \frac{\pi}{\omega},
\end{eqnarray*}

\noindent which is exactly one tick of her/his or her proper time. The subsequent
sound waves sent at each tick of the respective wrist watches of the
lower and upper observers reach the middle observer then just in time
one tick later.

From the perspective of the spheronic observer the situation looks
much simpler. For her or him, the lower soundwave was emitted at
\begin{eqnarray*}
z^{'} & = & \frac{1}{\gamma_{s}}\left(-\frac{\pi}{k}\frac{1}{\gamma_{s}}-v\left(-\frac{1}{\gamma_{s}}\frac{v}{c_{s}^{2}}\frac{\pi}{k}\right)\right),\\
 & = & \frac{\pi}{k}\frac{1}{\gamma_{s}^{2}}\left(-1+\frac{v^{2}}{c_{s}^{2}}\right),\\
 & = & -\frac{\pi}{k},
\end{eqnarray*}
which is simply the end of her/his ``lower'' ruler. The upper sound wave
was emitted at 
\begin{eqnarray*}
z^{'} & = & \frac{1}{\gamma_{s}}\left(\frac{\pi}{k}\frac{1}{\gamma_{s}}-v\left(\frac{1}{\gamma_{s}}\frac{v}{c_{s}^{2}}\frac{\pi}{k}\right)\right)\\
 & = & \frac{\pi}{k}\frac{1}{\gamma_{s}^{2}}\left(1-\frac{v^{2}}{c_{s}^{2}}\right),\\
 & = & \frac{\pi}{k},
\end{eqnarray*}
which is the end of the upper ruler. Both waves were emitted at
$\tau=0$ and are registered after one tick of proper time, the
spheronic observer calculates the speed of sound as $\Delta
z^{'}/\Delta t^{'}=c_{s}$ which is the same in both directions. There
is no chance for the spheronic observer to determine, with the
spheronic tools of measurement available in the spheronic universe,
her/his speed against the gas. For the spheronic observer there is no
preferred frame.  Using her or his spheronic tools and with the help
of comoving, synchronized fellow spheronic observers, he/she will be
able to set up a spheronic spacetime and he/she will eventually arrive
at a spheronic theory of relativity as an accurate reflection of the
spheronic observer's view (perception respectively) of the world he or
she inhabits.

This same situation appears differently for the external human
scientist. For such an observer, Newtonian (absolute) space and time
are the natural frame to work in, the gas is a physically measurable
entity and applying Newtonian mechanics to the gas results in the wave
equation. The perspective of the external human scientist looking at
the gas is the classical perspective of physics and as such also the
perspective of a 19th century physicist reasoning about an
ether-theory.  Starting with Poincare and Minkowski, many renowned
physicist have claimed that an ether theory with contraction and
dilation is mathematically equivalent to a theory of special
relativity

The difference between the ether theory and spherons in a gas lies in
the idea of matter. Classical matter is ``solid'' and ``rigid'', which
are ideas derived from our everyday experience with matter.  Spherons
in contrast are waves but their properties are in many respects close
to the properties attributed to real matter by modern physics, even
including the ``spheronic wave-particle dualism''
(Sec.~\ref{sec:OpMeth}). {\it It may thus be argued that contraction and
time dilation are no presents from above but that they are rather
simple consequences of the wave-nature of matter}. Whatever other
useful insights this may lead to, this can be exploited to establish
the existence of a spheronic theory of relativity.

This leads to the question of the didactic value of the present
approach. Explicit calculations such as above are instructive to get a
feeling for the ``seemingly strange'' properties of relativity. While
these calculations may as well be done within special relativity, the
spheronic idealized universe allows for the establishment of a flat
4-dimensional pseudo-Riemannian manifold right in front of our eyes
within a completely classical framework. Looking at something
happening within this idealized universe, it is most interesting to
change the perspective from the external human scientist (who may be
studying the propagation of created spherons in a large volume of
ideal gas), to an internal spheronic observer and back and reason about
the resulting changes.  Of course, in the real world, we are in the
role of the spheronic observer and the perspective of the human
scientist is not available to us.

Last but not least, the term ``Lorentz-composition'' has been chosen
here to avoid confusion with the notions accompanying the coordinate
transformations of special relativity. But it may be reasoned that
this is definitely more than a mere mathematical ``trick'' or a
meaningless curiosity, it definitely does have a physical meaning. The
above $\rho$ is a ``propagating standing wave'' and it is a fully
valid solution of the wave equation in its own right which is not
commonly found in textbooks. It can be used as an Ansatz in numerous
fields of physics where ever the wave equation plays a role, ranging
from classical mechanics, as is the case here, to gravitational waves
in General Relativity in the linear limit for weak fields.

\end{appendix}

% UNCOMMENT THE LINES BELOW IF YOU WISH TO USE BIBTEX
%\bibliographystyle{apj}
%\bibliography{yourbibfile}

\begin{thebibliography}{}
% References are listed as in the following example, for more examples, please
% see the PASA Style Guide

%\bibitem[Bell(1976)]{Bell76}Bell, J.~S., 1976,
%Progress in Scientific Culture, Vol. 1, No. 2 

\bibitem[de Broglie(1925)] {key-2}de Broglie, Louis (1925): Recherches
  sur la th\'eorie des Quanta, Annales de Physique, 10-\`eme s\'erie,
  p.22--128

%\bibitem[Feynman(1965)] {key-3}Feynman, Richard P. (1965): ``The Feynman lectures
%on physics - Volume III Quantum Mechanics'', BILINGUA Ausgabe, R.
%Oldenbourg und Addison-Wesley, München 1971

\bibitem[Minkowski(1908)] {Minkowski(1908)}Minkowski, Herrmann (1908): ``Raum und Zeit'', in
Jahresberichte der Deutschen Mathematiker-Vereinigung, B.G. Teubner,
Leipzig, 1909. 

\bibitem[Penrose(2005)] {Penrose(2005)}Penrose, Roger (2005): ``The Road to
Reality - A Complete Guide to the Laws of the Universe'', Vintage Books, London

\bibitem[Skudrzyk] {key-6}Skudrzyk, Eugen (1971): ``The Foundations of Acoustics
- Basic Mathematics and Basic Acoustics'', Springer-Verlag, Wien

\bibitem[Tipler 
\& Llewellyn(2002)]{Tipler02} Tipler, P.~A., \& Llewellyn, R.\ 2002, Modern Physics, 4th edition, by Paul A.~Tipler and Ralph Llewellyn.~W.~H.~Freeman Publishers, ISBN 0-71674345-0, 700pp


%\bibitem[Weinberg(666)]{key-7}Weinberg: zu ergänzen

\end{thebibliography}

\end{document}